\documentclass[prl,twocolumn,floatfix,superscriptaddress,longbibliography,nofootinbib]{revtex4-1}
\RequirePackage[sort&compress]{natbib}

\usepackage{amssymb,amsfonts,amsmath}
\usepackage{bm}
\usepackage[pdftex]{graphicx}
\usepackage{xcolor}
\usepackage{setspace}
\usepackage{subfigure}
\usepackage{comment}

\usepackage{hyperref}
\hypersetup{
    colorlinks=true,
    linkcolor=blue,
    filecolor=magenta,      
    urlcolor=blue,
}
 
\newcommand{\rev}[1]{\textcolor{black}{#1}}
\newcommand{\heading}[1]{\vspace{0.25truecm}\hspace{0.25truecm}\emph{#1.} }

\begin{document}

\title{Diffusion geometry unravels the emergence of functional clusters in collective phenomena}

\author{Manlio De Domenico\\
\normalsize{Departament d'Enginyeria Inform\`atica i Matem\`atiques, Universitat Rovira i Virgili, 43007 Tarragona, Spain}
}

\begin{abstract}
Collective phenomena emerge from the interaction of natural or artificial units with a complex organization. The interplay between structural patterns and dynamics might induce functional clusters that, in general, are different from topological ones. In biological systems, like the human brain, the overall functionality is often favored by the interplay between connectivity and synchronization dynamics, with functional clusters that do not coincide with anatomical modules in most cases. In social, socio-technical and engineering systems, the quest for consensus favors the emergence of clusters. 

Despite the unquestionable evidence for mesoscale organization of many complex systems and the heterogeneity of their inter-connectivity, a way to predict and identify the emergence of functional modules in collective phenomena continues to elude us. \rev{Here, we propose an approach based on random walk dynamics to define the diffusion distance between any pair of units in a networked system. Such a metric allows to exploit the underlying diffusion geometry} to provide a unifying framework for the intimate relationship between metastable synchronization, consensus and random search dynamics in complex networks, pinpointing the functional mesoscale organization of synthetic and biological systems.
\end{abstract}

\maketitle

The absence of a central authority coordinating the interactions among units of a complex system might lead to interesting collective phenomena, such as synchronization~\cite{arenas2008synchronization} in biological systems or consensus~\cite{olfati2007consensus} in social and technological networks. This type of self-organization is affected by the underlying structure, which for a wide variety of real systems is highly heterogenous~\cite{barabasi1999emergence} and modular~\cite{krause2003compartments,guimera2005functional}. Understanding the interplay between structure and dynamics of such systems has been, and still is, a major challenge in the study of complex systems. Empirical observations, confirmed by numerical simulations and theoretical predictions, suggest that complex systems with hierarchical and/or modular mesoscale organization of their units~\cite{fortunato2010community} are characterized by topological scales~\cite{arenas2006synchronization} and the emergence of functional clusters that might be, in general, different from topological ones. 

In this letter, we show that such functional clusters might be predicted and identified for a wide variety of complex networks. More specifically, for biological systems which can be modeled as networks of oscillators, and for systems of individuals or sensors attempting to reach consensus. The unifying picture is provided by diffusion geometry~\cite{coifman2005geometric}, developed one decade ago for nonlinear dimensionality reduction of complex data. This approach uses Markov processes to integrate local similarities at different scales, allowing to approximate the manifold which better describes the data while preserving their topological features. From a physical perspective, this approach relies on topological information gathered by random searches across time, a principle that has been used successfully in network science to unravel the topological mesoscale organization of a system based on how information flows through its units~\cite{rosvall2008maps,delvenne2010stability,schaub2012markov,della2013profiling,lambiotte2014markov,rosvall2014memory}.

\begin{figure}[!t]
\centering
\includegraphics[width=8.cm]{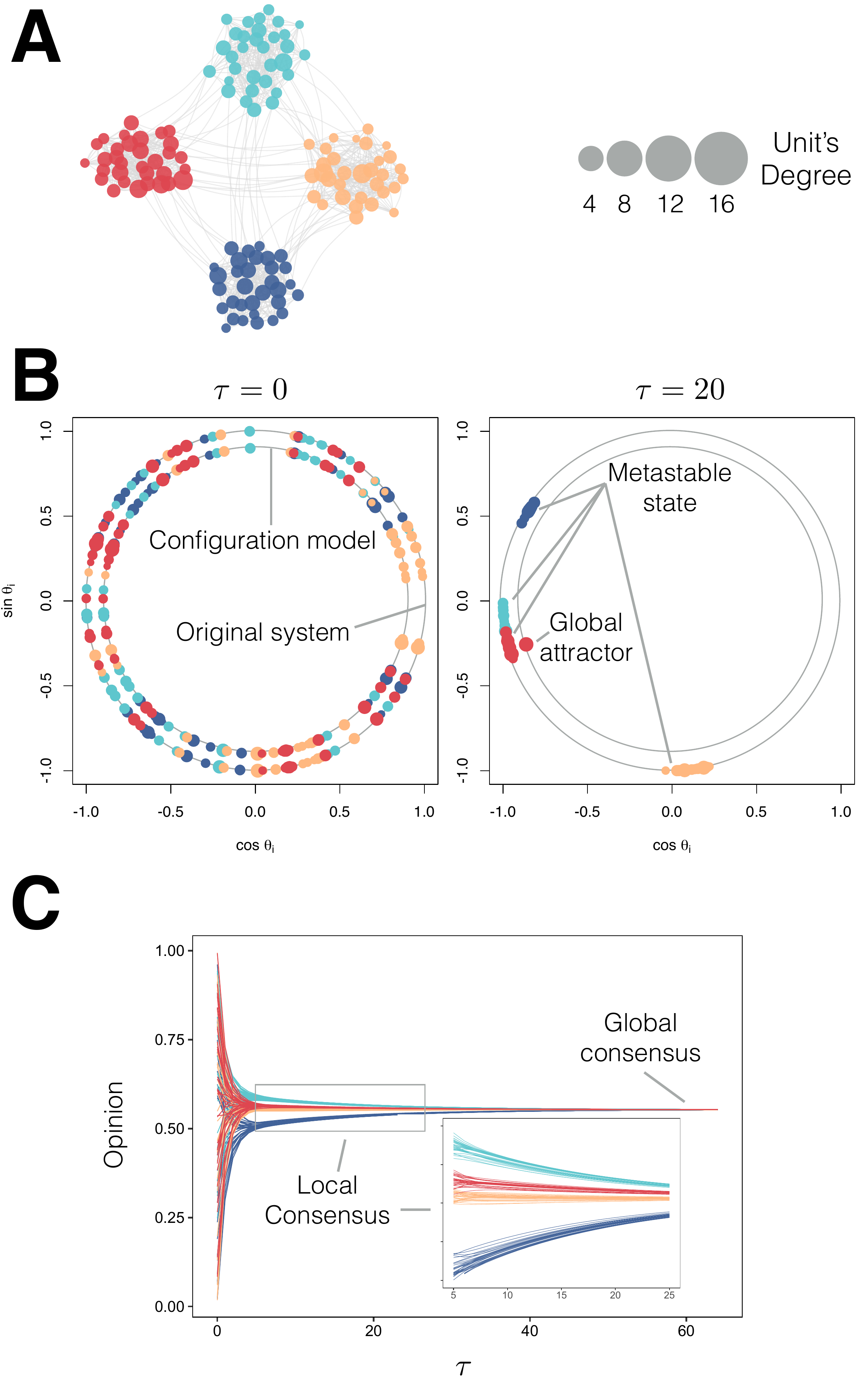}
\caption{\label{fig:synchro}\small{\textbf{Emergence of functional clusters.} (A) A Girvan-Newman benchmark network~\cite{newman2012communities} of $N=128$ oscillators. The mesoscale structure is organized into four clusters. (B) Phases $\theta_{i}$ ($i=1,2,...,N$) of identical oscillators ($\omega_{i}=0$) reported on a polar coordinate system with unitary radius for the original system (outer ring) and with smaller radius (inner ring) for one realization of its configuration model (which preserves the connectivity distribution of the original data and remove other correlations). The oscillators have initial phases uniformly distributed in $[0,2\pi]$ at time $\tau=0$ (left panel). They are free to interact each other, according to the underlying topology, and drive the systems to collective synchronization at $\tau=20$ (right panel). The original system reaches a metastable state -- with intra-cluster units synchronized to a common phase, with small fluctuations around a reference value -- whereas the configuration model -- where the mesoscale structure has been destroyed -- quickly reaches the global attractor. \rev{(C) Opinion-formation dynamics of agents in the DeGroot model of decentralized consensus~\cite{degroot1974reaching}. Each line represents the evolution of an opinion $x_{i}(t)$. In the metastable state local consensus is firstly achieved within clusters (see the inset) and later evolves into a collective opinion.}} }
\end{figure}

\heading{Synchronization dynamics} Let us indicate with $A_{ij}$ the entries of the adjacency matrix $\mathbf{A}$ representing the connections among a set of $N$ units (note that $A_{ij}=1$ if two units are connected and zero otherwise), each one encoding an oscillator with natural frequency $\omega_{i}$ and phase $\theta_{i}$. The dynamics of this networked system of oscillators has been widely studied in the last decades~\cite{arenas2008synchronization} and it is generally described by the Kuramoto model:
\begin{eqnarray}
\dot{\theta}_{i}(\tau) = \omega_{i} + \sum_{j=1}^{N} \sigma_{ij} A_{ij}\sin(\theta_{j}(\tau)-\theta_{i}(\tau)).
\end{eqnarray}
The choice of $\sigma_{ij}$, the mixing rate, determines the speed of convergence to a synchronized state, if any, and the behavior of the system in the thermodynamic limit $N\longrightarrow \infty$. It has been shown that, at variance with one's naive expectation, synchronizability does not necessarily correlate with the average distance between oscillators, which might be extraordinarily small in the case of strongly heterogeneous connectivity~\cite{cohen2003scale}. Such an heterogeneity might, in fact, suppress synchronization in networked oscillators which are coupled symmetrically with uniform coupling strength~\cite{nishikawa2003heterogeneity}. A solution to this apparent paradox~\cite{motter2005network} -- undermining the relevance of scale-free paradigm as a universal property of robust self-organizing phenomena favored by evolutionary dynamics~\cite{barabasi1999emergence} -- is to consider a mixing rate which is inversely proportional to node's degree $k_{i}=\sum\limits_{j}A_{ij}$, i.e. $\sigma_{ij}=K/k_{i}$, being $K$ an overall coupling constant (that we set equal to 1 in our analysis). This choice effectively reduces the dephasing effects in hubs, putting in a closer relationship the dynamics of synchronization close to the global attractor with the dynamics of information diffusion in the network, confirming that synchronizability does not only spread along shortest paths between two units but along all possible ones.

\begin{figure*}[!t]
\centering
\includegraphics[width=16cm]{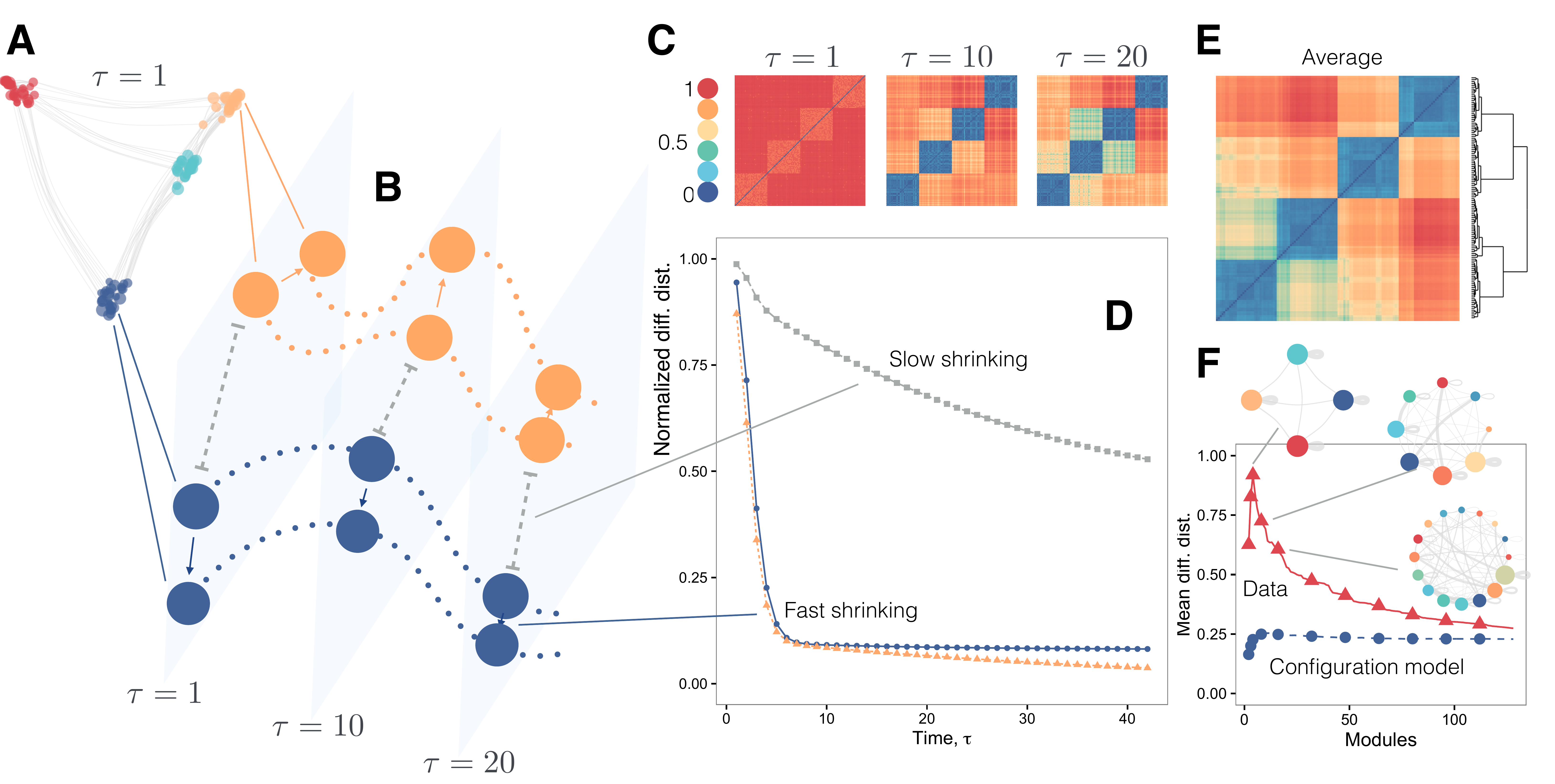}
\caption{\label{fig:meso}\small{\textbf{Identifying functional clusters in diffusion space.} (A) A Girvan-Newman benchmark network~\cite{newman2012communities} with four clusters, embedded in the Euclidean space by using multidimensional scaling applied to the diffusion-distance matrix $\boldsymbol{\Delta}_{\tau=1}$. (B) Two units from the same cluster are closer across time ($\tau$) than units from different clusters. (C) Diffusion-distance matrices corresponding to different times: the mesoscale structure becomes more evident as time goes by. (D) The diffusion-distance matrix at time $\tau$ is normalized as $\boldsymbol{\Delta}_{\tau}/\max\limits_{ij}(\Delta_{ij}(\tau))$ ($i,j=1,2,...,N$): the normalized distance between units from the same cluster quickly shrinks, while the one between units from different clusters slowly shrinks; this peculiar behavior is used to probe the mesoscale structure at different resolutions. (E) The diffusion-distance matrix built by averaging the matrices up to a certain time $\tau_{max}$ accounts for persistence of mesoscale across time and it is used to unveil its hierarchical organization by means of hierarchical clustering. (F) All resulting hierarchies are screened and the corresponding networks of clusters are built. The network which better represents the mesoscale structure is the one where the average diffusion distance among clusters is maximized. The significance of such a structure can be easily quantified by comparing with the result obtained from a network model preserving the degree distribution of the original data while destroying other correlations (i.e., a configuration model). See the Supplementary Information for further analysis of synthetic networks.}}
\end{figure*}

In complex networks with a well defined mesoscale organization, nodes belonging to the same cluster tend to synchronize to a common phase, not necessarily equal for all clusters, while the dynamics towards synchronization evolves~\cite{oh2005modular}. If the natural frequency is the same for all units, there is only one attractor for the dynamics, corresponding to the point where all phases are the same, i.e., $\theta_{i}(\tau \longrightarrow\infty)=\theta^{\star}$ for $i=1,2,...,N$ and $\tau$ representing time. Numerical experiments show that a strong cluster organization favors a metastable synchronized state, where $\theta_{i}\simeq \theta_{j}$ if nodes $i$ and $j$ belong to the same cluster. 
In this peculiar state -- and for a sufficiently small amount of time -- contributions from units which act as bridges with other clusters might be neglected with respect to the larger number of intra-cluster contributions. The overall dynamics therefore consists of a first phase, where intra-cluster synchronization takes place, followed by a second phase where cluster-cluster synchronization emerges, slowly driving the system towards its global attractor (see Fig.~\ref{fig:synchro}). During both phases, $\sin(\theta_{j}-\theta_{i})\simeq (\theta_{j}-\theta_{i})$ and the dynamics can be approximately described by 
\begin{eqnarray}
\label{eq:laplacian}
\dot{\boldsymbol{\theta}} = -\mathbf{\tilde{L}} \boldsymbol{\theta},
\end{eqnarray}
where $\mathbf{\tilde{L}}=\mathbf{I}-\mathbf{D}^{-1}\mathbf{A}$ is the normalized Laplacian matrix, $\mathbf{I}$ is the identity matrix, $D_{ii}=k_{i}$ and $D_{ij}=0$ for $i\neq j$. The matrix $\mathbf{\tilde{L}}$ governing the dynamics is the same which governs the diffusion of a random walker and the probability to find it in a certain node at a certain time step, as we will see later. During the metastable state, we can describe the common phase of nodes which are clustered together by $\theta_{0}^{C_{m}}$, with $m=1,2,...,M$ indicating the cluster, and we indicate with $\boldsymbol{\theta}_{0}$ the vector $(\theta_{0}^{C_{1}}, \theta_{0}^{C_{2}}, ..., \theta_{0}^{C_{M}})$. Let us introduce the rectangular matrix $\mathbf{S}$ encoding the (unknown) mesoscale organization of the system, i.e., $S_{im}=1$ if node $i$ belongs to cluster $m$ and it is zero otherwise. Such definitions allow us to write the state vector in a very compact form as $\mathbf{z} = \mathbf{S}\boldsymbol{\theta}_{0}$. Let us make a localized small perturbation on the phase of unit $i$: the perturbed state can be written as $\mathbf{z}_{i}=\mathbf{z} + \delta \theta_{0}\mathbf{v}_{i}$, being $\mathbf{v}_{i}$ the canonical vector with $i-$th component equal to 1 and $\delta \theta_{0}\ll1$. By assuming the metastable state as the initial condition, the state of the system at time $\tau$ is given by $\boldsymbol{\theta}(\tau;i)=\exp{(-\tau\mathbf{\tilde{L}})}\mathbf{z}_{i}$. It is plausible to expect that the magnitude of the difference between the evolution of the perturbed states $\mathbf{z}_{i}$ and $\mathbf{z}_{j}$ is small when the corresponding nodes belong to the same cluster and larger when this is not the case. \rev{We define the synchronizability distance between two nodes by
\begin{eqnarray}
s^{2}_{\tau}(i,j)=[\boldsymbol{\tilde{\theta}}(\tau;i)-\boldsymbol{\tilde{\theta}}(\tau;j)]^{2},
\end{eqnarray}
with $\boldsymbol{\tilde{\theta}}(\tau;i)=\mathbf{z}_{i}\exp(-\tau\mathbf{\tilde{L}})$, to quantify how easy for two nodes is to reach a common phase during a metastable state. 
Intriguingly, the synchronizability distance reduces to $s^{2}_{\tau}(i,j)\propto\left[ (\mathbf{v}_{i}-\mathbf{v}_{j})e^{-\tau\mathbf{\tilde{L}}} \right]^{2}$, where the right-hand side is better known as diffusion distance~\cite{belkin2001laplacian}.}

\heading{Consensus dynamics} In a social context, as well as in a system of sensors, decision-making processes require individuals (or units) to exchange information to self-organize and, under certain circumstances -- such as the absence of coordinating authorities or external influences -- the emergence of consensus is observed~\cite{olfati2004consensus,olfati2007consensus}. A distributed consensus dynamics based on a linear protocol exists and it is governed by the Laplacian matrix of the network. 
Because of the natural heterogeneity observed in this type of systems~\cite{liljeros2001web}, it is desirable to define a consensus dynamics where the weight due to high connectivity of a few individuals is somehow compensated, for instance by rescaling the amount of exchanged information by their degree. \rev{This type of decentralized opinion-formation dynamics is equivalent to a continuous-time DeGroot model~\cite{degroot1974reaching} and can mathematically described as in Eq.~(\ref{eq:laplacian}), with the opinion vector} $\mathbf{x}(\tau)$ playing the role of the phase vector $\boldsymbol{\theta}(\tau)$. It is straightforward to show that the weighted-average consensus is asymptotically reached~\cite{olfati2007consensus}. Similarly to the case of synchronization, we expect that in a network with a mesoscale organization, individuals or units within a cluster tend to reach consensus before, successively driving the collective dynamics of the system towards the overall consensus \rev{(see Fig.~\ref{fig:synchro}C)}. To better understand this process, we consider that the system is in a consensus state except for node $i$, e.g. $\mathbf{x}(0)=\mathbf{v}_{i}$. We consider the same setup with another node $j\neq i$ and then we track the evolution of both states over time. \rev{We introduce the consensus distance
\begin{eqnarray}
c^{2}_{\tau}(i,j)=[\mathbf{\tilde{x}}(\tau;i)-\mathbf{\tilde{x}}(\tau;j)]^{2},
\end{eqnarray}
with $\mathbf{\tilde{x}}(\tau;i)=\mathbf{v}_{i}\exp(-\tau\mathbf{\tilde{L}})$, under the plausible assumption that, like in the case of synchronization, this distance tends to be small if the two nodes belong to the same cluster and it is larger otherwise. This distance can be rewritten as $c^{2}_{\tau}(i,j)=\left[ (\mathbf{v}_{i}-\mathbf{v}_{j}) e^{-\tau\mathbf{\tilde{L}}} \right]^{2}$, where the right-hand side is the diffusion distance.}

\heading{Using diffusion geometry to reveal functional clusters} 
The dynamics describing how a piece of information diffuses through networked systems has been well studied for classical~\cite{noh2004random} and multilayer networks~\cite{dedomenico2014navigability,de2016physics} (\rev{see Ref.~\cite{masuda2016random} for a thorough review}). The probability to find the random walker in any node after a certain amount of time $\tau$ is given by the solution of the master equation
\begin{eqnarray}
\dot{\mathbf{p}}(\tau)=-\mathbf{p}(\tau)\mathbf{\tilde{L}},
\end{eqnarray}
where $\mathbf{\tilde{L}}$ is the normalized Laplacian matrix we have discussed before. The general solution is given by $\mathbf{p}(\tau)=\mathbf{p}(0)\exp{(-\tau \mathbf{\tilde{L}})}$. Here, we indicate by $\mathbf{p}(\tau|i)=\mathbf{v}_{i}\exp{(-\tau \mathbf{\tilde{L}})}$ the probability vector corresponding to the initial condition where the walker's origin is in node $i$ with probability 1 (i.e., $\mathbf{p}(0)=\mathbf{v}_{i}$).

We exploit the intriguing connection between the measure of synchronizability in the metastable state, consensus and information diffusion to identify synchronization/consensus clusters, after mapping this problem into a hidden geometric space induced by Markov dynamics. The diffusion distance~\cite{belkin2001laplacian} between nodes $i$ and $j$ is defined by
\begin{eqnarray}
d^{2}_{\tau}(i,j)=\left[\mathbf{p}(\tau|i) - \mathbf{p}(\tau|j)\right]^{2},
\end{eqnarray}
where $p_{k}(\tau|i)$ encodes the probability to find a random walker originated in $i$ at node $k$, at time $\tau$. Diffusion maps, built on this concept, are widely adopted for low-dimensional embedding of high-dimensional data~\cite{coifman2005geometric,jones2008manifold} and provide a unified probabilistic interpretation for spectral embedding and clustering algorithms~\cite{nadler2008diffusion}, among others. The diffusion distance between two nodes is small if there are many paths which connect them, allowing information to be easily exchanged. We can exploit this property to gather insight about physical processes, such as information diffusion, and collective phenomena with emergent behavior, such as synchronization and consensus dynamics. In fact, in a complex network where units are organized in functional clusters, the diffusion distance among nodes belonging to the same cluster must be small, because the mesoscale structure favors the information exchange within the clusters rather than across them. The relationships among these processes is made explicit by the identities $s^{2}_{\tau}=\delta\theta_{0}d^{2}_{\tau}(i,j)$ and $c^{2}_{\tau}(i,j)=d^{2}_{\tau}(i,j)$.

At a specific time delay $\tau$, the diffusion distances among all pair of nodes define a matrix $\boldsymbol{\Delta}_{\tau}$, that we name diffusion-distance matrix in the following. To obtain a geometrical intuition about its meaning, we can embed the units into a low-dimensional Euclidean space by using, for instance, multidimensional scaling (Fig.~\ref{fig:meso}A). In this diffusion space, closer points correspond to units with smaller diffusion distance, i.e., to nodes that successfully exchange information in less than $\tau$ steps (Fig.~\ref{fig:meso}B). Important consequences of this approach include the mapping from network's mesoscale to clusters in space (Fig.~\ref{fig:meso}C) and the identification of hierarchies at multiple resolutions. When $\tau$ is small, micro scale structure is revealed, while for increasing $\tau$ the mesoscale is screened until the macro scale structure is captured.

For specific applications, it might be useful to identify the mesoscale structure which provides the best coarse-groaning of the system, with respect to certain criteria. We use the persistence of the mesoscale across time, if any, to characterize the system. By construction, the diffusion distance between two units tends to zero for increasing time, it is therefore necessary to normalize it appropriately to allow the comparison between the cluster formation at different values of $\tau$. As shown in Fig.~\ref{fig:meso}D, this can be accomplished by using the normalized matrix $\boldsymbol{\tilde{\Delta}}_{\tau}=\boldsymbol{\Delta}_{\tau}/\max\limits_{ij}(\Delta_{ij}(\tau))$, with the persistence of clusters being encoded in the persistence of the diffusion distance between their units. We exploit the fact that the normalized diffusion distance quickly shrinks for intra-cluster nodes, to guarantee that the average diffusion-distance matrix, defined by $\boldsymbol{\bar{\Delta}} = \tau_{max}^{-1}\sum\limits_{\tau=1}^{\tau_{max}}\boldsymbol{\Delta}_{\tau}$ -- where $\tau_{max}$ is a temporal cutoff -- will preserve this geometrical persistence. For $\tau_{max}\approx N$, i.e., the size of the system, the results obtained from the matrix $\boldsymbol{\bar{\Delta}}$ are robust to the choice of this cutoff. It $\tau_{max}\ll N$, the random walkers have not enough time to search through the system, and only the mesoscale closer to the micro scale can be revealed. Conversely, if $\tau_{max}\gg N$, the information gathered during the search is washed out and only the macro scale can be captured. The hierarchical clustering of units in the diffusion space of average distances reveals the most persistent clusters and their hierarchical organization (Fig.~\ref{fig:meso}E). To understand which hierarchy better represents the mesoscale structure, it is natural to analyze the corresponding network of clusters, where each node is a functional super-unit -- consisting of units belonging to the same functional cluster -- and connections between super-units are weighted by inter-cluster connectivity. The average diffusion distance among super-units is expected to be maximum when diffusion between clusters is extremely hindered; this happens when the most representative functional mesoscale is captured, and it is significantly different from random expectation (Fig.~\ref{fig:meso}F).

To better understand the relationship between structural communities, due to purely topological connectivity, and the functional clusters, due to the interplay between structure and dynamics previously described, we have generated and analyzed ensembles of Girvan-Newman networks~\cite{girvan2002community}, while varying the ratio between inter- and intra-community connectivity. Diffusion geometry identifies clusters in agreement with structural ones when this ratio is very small -- i.e., when the structural mesoscale is strongly organized into well-defined clusters -- and provides different results for larger ratios, by identifying a larger number of functional modules, compared to other methods~\cite{reichardt2004detecting,blondel2008fast,rosvall2007information,rosvall2008maps} (see Suppl. Fig.~3).

Given the expected difference between topological and functional clusters, as an application of our framework we analyze an empirical network providing anatomical connectivity within and between visual cortical and sensorimotor areas in Macaque brain~\cite{negyessy2006prediction}. Our analysis (see Suppl. Fig.~4) reveals a hierarchical functional organization of cortical units, significantly different from what should be expected from a network with the same connectivity distribution in absence of correlations. The importance of ventral intraparietal (VIP) region in bridging the two functional areas is manifested from the analysis, in perfect agreement with previous findings~\cite{negyessy2006prediction}. Other key functional modules, such as
areas 46 and 7a, are successfully identified, confirming studies based on neural collective behavior measured from transfer entropy functional connectivity and blood oxygenation level-dependent correlation patterns~\cite{honey2007network}. It is worth remarking that despite our results are not based on external functional information, they provide results comparable with existing knowledge obtained from that information. \rev{The analysis of similarities among the identified functional clusters, the anatomical ones and the structural mesoscale organization obtained from the spin-glass approach~\cite{reichardt2004detecting}, shows that our diffusion geometry framework identifies a functional organization that is distinct from the structural one (see Suppl. Fig.~5).}

As diffusion mapping revolutionized applied math and machine learning, we envision many potential applications in complex systems physics based on the unifying framework of diffusion geometry. Complementary to approaches based on network's hidden geometry deduced from structural properties~\cite{serrano2008self,boguna2009navigability,papadopoulos2012popularity,kleineberg2016hidden}, future applications to multilayer networks~\cite{de2013mathematical,del2016synchronization,de2016physics} will allow to gain further insight on collective phenomena emerging from the interplay between structure and dynamics in such systems.


\vspace{0.25truecm}
\begin{acknowledgments}
\emph{The author thanks Alex Arenas, Joan T. Matamalas and Massimo Stella for fruitful discussions. MDD acknowledges financial support from MINECO program Juan de la Cierva (IJCI-2014-20225). }
\end{acknowledgments}

\bibliographystyle{apsrev4-1}
\bibliography{biblio}

\begin{thebibliography}{41}%
\makeatletter
\providecommand \@ifxundefined [1]{%
 \@ifx{#1\undefined}
}%
\providecommand \@ifnum [1]{%
 \ifnum #1\expandafter \@firstoftwo
 \else \expandafter \@secondoftwo
 \fi
}%
\providecommand \@ifx [1]{%
 \ifx #1\expandafter \@firstoftwo
 \else \expandafter \@secondoftwo
 \fi
}%
\providecommand \natexlab [1]{#1}%
\providecommand \enquote  [1]{``#1''}%
\providecommand \bibnamefont  [1]{#1}%
\providecommand \bibfnamefont [1]{#1}%
\providecommand \citenamefont [1]{#1}%
\providecommand \href@noop [0]{\@secondoftwo}%
\providecommand \href [0]{\begingroup \@sanitize@url \@href}%
\providecommand \@href[1]{\@@startlink{#1}\@@href}%
\providecommand \@@href[1]{\endgroup#1\@@endlink}%
\providecommand \@sanitize@url [0]{\catcode `\\12\catcode `\$12\catcode
  `\&12\catcode `\#12\catcode `\^12\catcode `\_12\catcode `\%12\relax}%
\providecommand \@@startlink[1]{}%
\providecommand \@@endlink[0]{}%
\providecommand \url  [0]{\begingroup\@sanitize@url \@url }%
\providecommand \@url [1]{\endgroup\@href {#1}{\urlprefix }}%
\providecommand \urlprefix  [0]{URL }%
\providecommand \Eprint [0]{\href }%
\providecommand \doibase [0]{http://dx.doi.org/}%
\providecommand \selectlanguage [0]{\@gobble}%
\providecommand \bibinfo  [0]{\@secondoftwo}%
\providecommand \bibfield  [0]{\@secondoftwo}%
\providecommand \translation [1]{[#1]}%
\providecommand \BibitemOpen [0]{}%
\providecommand \bibitemStop [0]{}%
\providecommand \bibitemNoStop [0]{.\EOS\space}%
\providecommand \EOS [0]{\spacefactor3000\relax}%
\providecommand \BibitemShut  [1]{\csname bibitem#1\endcsname}%
\let\auto@bib@innerbib\@empty
\bibitem [{\citenamefont {Arenas}\ \emph {et~al.}(2008)\citenamefont {Arenas},
  \citenamefont {D{\'\i}az-Guilera}, \citenamefont {Kurths}, \citenamefont
  {Moreno},\ and\ \citenamefont {Zhou}}]{arenas2008synchronization}%
  \BibitemOpen
  \bibfield  {author} {\bibinfo {author} {\bibfnamefont {A.}~\bibnamefont
  {Arenas}}, \bibinfo {author} {\bibfnamefont {A.}~\bibnamefont
  {D{\'\i}az-Guilera}}, \bibinfo {author} {\bibfnamefont {J.}~\bibnamefont
  {Kurths}}, \bibinfo {author} {\bibfnamefont {Y.}~\bibnamefont {Moreno}}, \
  and\ \bibinfo {author} {\bibfnamefont {C.}~\bibnamefont {Zhou}},\ }\href@noop
  {} {\bibfield  {journal} {\bibinfo  {journal} {Physics reports}\ }\textbf
  {\bibinfo {volume} {469}},\ \bibinfo {pages} {93} (\bibinfo {year}
  {2008})}\BibitemShut {NoStop}%
\bibitem [{\citenamefont {Olfati-Saber}\ \emph {et~al.}(2007)\citenamefont
  {Olfati-Saber}, \citenamefont {Fax},\ and\ \citenamefont
  {Murray}}]{olfati2007consensus}%
  \BibitemOpen
  \bibfield  {author} {\bibinfo {author} {\bibfnamefont {R.}~\bibnamefont
  {Olfati-Saber}}, \bibinfo {author} {\bibfnamefont {J.~A.}\ \bibnamefont
  {Fax}}, \ and\ \bibinfo {author} {\bibfnamefont {R.~M.}\ \bibnamefont
  {Murray}},\ }\href@noop {} {\bibfield  {journal} {\bibinfo  {journal}
  {Proceedings of the IEEE}\ }\textbf {\bibinfo {volume} {95}},\ \bibinfo
  {pages} {215} (\bibinfo {year} {2007})}\BibitemShut {NoStop}%
\bibitem [{\citenamefont {Barab{\'a}si}\ and\ \citenamefont
  {Albert}(1999)}]{barabasi1999emergence}%
  \BibitemOpen
  \bibfield  {author} {\bibinfo {author} {\bibfnamefont {A.}~\bibnamefont
  {Barab{\'a}si}}\ and\ \bibinfo {author} {\bibfnamefont {R.}~\bibnamefont
  {Albert}},\ }\href@noop {} {\bibfield  {journal} {\bibinfo  {journal}
  {Science}\ }\textbf {\bibinfo {volume} {286}},\ \bibinfo {pages} {509}
  (\bibinfo {year} {1999})}\BibitemShut {NoStop}%
\bibitem [{\citenamefont {Krause}\ \emph {et~al.}(2003)\citenamefont {Krause},
  \citenamefont {Frank}, \citenamefont {Mason}, \citenamefont {Ulanowicz},\
  and\ \citenamefont {Taylor}}]{krause2003compartments}%
  \BibitemOpen
  \bibfield  {author} {\bibinfo {author} {\bibfnamefont {A.~E.}\ \bibnamefont
  {Krause}}, \bibinfo {author} {\bibfnamefont {K.~A.}\ \bibnamefont {Frank}},
  \bibinfo {author} {\bibfnamefont {D.~M.}\ \bibnamefont {Mason}}, \bibinfo
  {author} {\bibfnamefont {R.~E.}\ \bibnamefont {Ulanowicz}}, \ and\ \bibinfo
  {author} {\bibfnamefont {W.~W.}\ \bibnamefont {Taylor}},\ }\href@noop {}
  {\bibfield  {journal} {\bibinfo  {journal} {Nature}\ }\textbf {\bibinfo
  {volume} {426}},\ \bibinfo {pages} {282} (\bibinfo {year}
  {2003})}\BibitemShut {NoStop}%
\bibitem [{\citenamefont {Guimera}\ and\ \citenamefont
  {Amaral}(2005)}]{guimera2005functional}%
  \BibitemOpen
  \bibfield  {author} {\bibinfo {author} {\bibfnamefont {R.}~\bibnamefont
  {Guimera}}\ and\ \bibinfo {author} {\bibfnamefont {L.~A.~N.}\ \bibnamefont
  {Amaral}},\ }\href@noop {} {\bibfield  {journal} {\bibinfo  {journal}
  {Nature}\ }\textbf {\bibinfo {volume} {433}},\ \bibinfo {pages} {895}
  (\bibinfo {year} {2005})}\BibitemShut {NoStop}%
\bibitem [{\citenamefont {Fortunato}(2010)}]{fortunato2010community}%
  \BibitemOpen
  \bibfield  {author} {\bibinfo {author} {\bibfnamefont {S.}~\bibnamefont
  {Fortunato}},\ }\href@noop {} {\bibfield  {journal} {\bibinfo  {journal}
  {Physics reports}\ }\textbf {\bibinfo {volume} {486}},\ \bibinfo {pages} {75}
  (\bibinfo {year} {2010})}\BibitemShut {NoStop}%
\bibitem [{\citenamefont {Arenas}\ \emph {et~al.}(2006)\citenamefont {Arenas},
  \citenamefont {D{\'\i}az-Guilera},\ and\ \citenamefont
  {P{\'e}rez-Vicente}}]{arenas2006synchronization}%
  \BibitemOpen
  \bibfield  {author} {\bibinfo {author} {\bibfnamefont {A.}~\bibnamefont
  {Arenas}}, \bibinfo {author} {\bibfnamefont {A.}~\bibnamefont
  {D{\'\i}az-Guilera}}, \ and\ \bibinfo {author} {\bibfnamefont {C.~J.}\
  \bibnamefont {P{\'e}rez-Vicente}},\ }\href@noop {} {\bibfield  {journal}
  {\bibinfo  {journal} {Physical Review Letters}\ }\textbf {\bibinfo {volume}
  {96}},\ \bibinfo {pages} {114102} (\bibinfo {year} {2006})}\BibitemShut
  {NoStop}%
\bibitem [{\citenamefont {Coifman}\ \emph {et~al.}(2005)\citenamefont
  {Coifman}, \citenamefont {Lafon}, \citenamefont {Lee}, \citenamefont
  {Maggioni}, \citenamefont {Nadler}, \citenamefont {Warner},\ and\
  \citenamefont {Zucker}}]{coifman2005geometric}%
  \BibitemOpen
  \bibfield  {author} {\bibinfo {author} {\bibfnamefont {R.~R.}\ \bibnamefont
  {Coifman}}, \bibinfo {author} {\bibfnamefont {S.}~\bibnamefont {Lafon}},
  \bibinfo {author} {\bibfnamefont {A.~B.}\ \bibnamefont {Lee}}, \bibinfo
  {author} {\bibfnamefont {M.}~\bibnamefont {Maggioni}}, \bibinfo {author}
  {\bibfnamefont {B.}~\bibnamefont {Nadler}}, \bibinfo {author} {\bibfnamefont
  {F.}~\bibnamefont {Warner}}, \ and\ \bibinfo {author} {\bibfnamefont {S.~W.}\
  \bibnamefont {Zucker}},\ }\href@noop {} {\bibfield  {journal} {\bibinfo
  {journal} {PNAS}\ }\textbf {\bibinfo {volume} {102}},\ \bibinfo {pages}
  {7426} (\bibinfo {year} {2005})}\BibitemShut {NoStop}%
\bibitem [{\citenamefont {Rosvall}\ and\ \citenamefont
  {Bergstrom}(2008)}]{rosvall2008maps}%
  \BibitemOpen
  \bibfield  {author} {\bibinfo {author} {\bibfnamefont {M.}~\bibnamefont
  {Rosvall}}\ and\ \bibinfo {author} {\bibfnamefont {C.~T.}\ \bibnamefont
  {Bergstrom}},\ }\href@noop {} {\bibfield  {journal} {\bibinfo  {journal}
  {PNAS}\ }\textbf {\bibinfo {volume} {105}},\ \bibinfo {pages} {1118}
  (\bibinfo {year} {2008})}\BibitemShut {NoStop}%
\bibitem [{\citenamefont {Delvenne}\ \emph {et~al.}(2010)\citenamefont
  {Delvenne}, \citenamefont {Yaliraki},\ and\ \citenamefont
  {Barahona}}]{delvenne2010stability}%
  \BibitemOpen
  \bibfield  {author} {\bibinfo {author} {\bibfnamefont {J.}~\bibnamefont
  {Delvenne}}, \bibinfo {author} {\bibfnamefont {S.}~\bibnamefont {Yaliraki}},
  \ and\ \bibinfo {author} {\bibfnamefont {M.}~\bibnamefont {Barahona}},\
  }\href@noop {} {\bibfield  {journal} {\bibinfo  {journal} {PNAS}\ }\textbf
  {\bibinfo {volume} {107}},\ \bibinfo {pages} {12755} (\bibinfo {year}
  {2010})}\BibitemShut {NoStop}%
\bibitem [{\citenamefont {Schaub}\ \emph {et~al.}(2012)\citenamefont {Schaub},
  \citenamefont {Delvenne}, \citenamefont {Yaliraki},\ and\ \citenamefont
  {Barahona}}]{schaub2012markov}%
  \BibitemOpen
  \bibfield  {author} {\bibinfo {author} {\bibfnamefont {M.~T.}\ \bibnamefont
  {Schaub}}, \bibinfo {author} {\bibfnamefont {J.-C.}\ \bibnamefont
  {Delvenne}}, \bibinfo {author} {\bibfnamefont {S.~N.}\ \bibnamefont
  {Yaliraki}}, \ and\ \bibinfo {author} {\bibfnamefont {M.}~\bibnamefont
  {Barahona}},\ }\href@noop {} {\bibfield  {journal} {\bibinfo  {journal} {PloS
  one}\ }\textbf {\bibinfo {volume} {7}},\ \bibinfo {pages} {e32210} (\bibinfo
  {year} {2012})}\BibitemShut {NoStop}%
\bibitem [{\citenamefont {Della~Rossa}\ \emph {et~al.}(2013)\citenamefont
  {Della~Rossa}, \citenamefont {Dercole},\ and\ \citenamefont
  {Piccardi}}]{della2013profiling}%
  \BibitemOpen
  \bibfield  {author} {\bibinfo {author} {\bibfnamefont {F.}~\bibnamefont
  {Della~Rossa}}, \bibinfo {author} {\bibfnamefont {F.}~\bibnamefont
  {Dercole}}, \ and\ \bibinfo {author} {\bibfnamefont {C.}~\bibnamefont
  {Piccardi}},\ }\href@noop {} {\bibfield  {journal} {\bibinfo  {journal}
  {Scientific reports}\ }\textbf {\bibinfo {volume} {3}},\ \bibinfo {pages}
  {1467} (\bibinfo {year} {2013})}\BibitemShut {NoStop}%
\bibitem [{\citenamefont {Lambiotte}\ \emph {et~al.}(2014)\citenamefont
  {Lambiotte}, \citenamefont {Delvenne},\ and\ \citenamefont
  {Barahona}}]{lambiotte2014markov}%
  \BibitemOpen
  \bibfield  {author} {\bibinfo {author} {\bibfnamefont {R.}~\bibnamefont
  {Lambiotte}}, \bibinfo {author} {\bibfnamefont {J.~C.}\ \bibnamefont
  {Delvenne}}, \ and\ \bibinfo {author} {\bibfnamefont {M.}~\bibnamefont
  {Barahona}},\ }\href@noop {} {\bibfield  {journal} {\bibinfo  {journal} {IEEE
  Transactions on Network Science and Engineering}\ }\textbf {\bibinfo {volume}
  {1}},\ \bibinfo {pages} {76} (\bibinfo {year} {2014})}\BibitemShut {NoStop}%
\bibitem [{\citenamefont {Rosvall}\ \emph {et~al.}(2014)\citenamefont
  {Rosvall}, \citenamefont {Esquivel}, \citenamefont {Lancichinetti},
  \citenamefont {West},\ and\ \citenamefont {Lambiotte}}]{rosvall2014memory}%
  \BibitemOpen
  \bibfield  {author} {\bibinfo {author} {\bibfnamefont {M.}~\bibnamefont
  {Rosvall}}, \bibinfo {author} {\bibfnamefont {A.~V.}\ \bibnamefont
  {Esquivel}}, \bibinfo {author} {\bibfnamefont {A.}~\bibnamefont
  {Lancichinetti}}, \bibinfo {author} {\bibfnamefont {J.~D.}\ \bibnamefont
  {West}}, \ and\ \bibinfo {author} {\bibfnamefont {R.}~\bibnamefont
  {Lambiotte}},\ }\href@noop {} {\bibfield  {journal} {\bibinfo  {journal}
  {Nature Communications}\ }\textbf {\bibinfo {volume} {5}},\ \bibinfo {pages}
  {4630} (\bibinfo {year} {2014})}\BibitemShut {NoStop}%
\bibitem [{\citenamefont {Newman}(2012)}]{newman2012communities}%
  \BibitemOpen
  \bibfield  {author} {\bibinfo {author} {\bibfnamefont {M.~E.}\ \bibnamefont
  {Newman}},\ }\href@noop {} {\bibfield  {journal} {\bibinfo  {journal} {Nature
  Physics}\ }\textbf {\bibinfo {volume} {8}},\ \bibinfo {pages} {25} (\bibinfo
  {year} {2012})}\BibitemShut {NoStop}%
\bibitem [{\citenamefont {DeGroot}(1974)}]{degroot1974reaching}%
  \BibitemOpen
  \bibfield  {author} {\bibinfo {author} {\bibfnamefont {M.~H.}\ \bibnamefont
  {DeGroot}},\ }\href@noop {} {\bibfield  {journal} {\bibinfo  {journal}
  {Journal of the American Statistical Association}\ }\textbf {\bibinfo
  {volume} {69}},\ \bibinfo {pages} {118} (\bibinfo {year} {1974})}\BibitemShut
  {NoStop}%
\bibitem [{\citenamefont {Cohen}\ and\ \citenamefont
  {Havlin}(2003)}]{cohen2003scale}%
  \BibitemOpen
  \bibfield  {author} {\bibinfo {author} {\bibfnamefont {R.}~\bibnamefont
  {Cohen}}\ and\ \bibinfo {author} {\bibfnamefont {S.}~\bibnamefont {Havlin}},\
  }\href@noop {} {\bibfield  {journal} {\bibinfo  {journal} {Physical Review
  Letters}\ }\textbf {\bibinfo {volume} {90}},\ \bibinfo {pages} {058701}
  (\bibinfo {year} {2003})}\BibitemShut {NoStop}%
\bibitem [{\citenamefont {Nishikawa}\ \emph {et~al.}(2003)\citenamefont
  {Nishikawa}, \citenamefont {Motter}, \citenamefont {Lai},\ and\ \citenamefont
  {Hoppensteadt}}]{nishikawa2003heterogeneity}%
  \BibitemOpen
  \bibfield  {author} {\bibinfo {author} {\bibfnamefont {T.}~\bibnamefont
  {Nishikawa}}, \bibinfo {author} {\bibfnamefont {A.~E.}\ \bibnamefont
  {Motter}}, \bibinfo {author} {\bibfnamefont {Y.-C.}\ \bibnamefont {Lai}}, \
  and\ \bibinfo {author} {\bibfnamefont {F.~C.}\ \bibnamefont {Hoppensteadt}},\
  }\href@noop {} {\bibfield  {journal} {\bibinfo  {journal} {Physical Review
  Letters}\ }\textbf {\bibinfo {volume} {91}},\ \bibinfo {pages} {014101}
  (\bibinfo {year} {2003})}\BibitemShut {NoStop}%
\bibitem [{\citenamefont {Motter}\ \emph {et~al.}(2005)\citenamefont {Motter},
  \citenamefont {Zhou},\ and\ \citenamefont {Kurths}}]{motter2005network}%
  \BibitemOpen
  \bibfield  {author} {\bibinfo {author} {\bibfnamefont {A.~E.}\ \bibnamefont
  {Motter}}, \bibinfo {author} {\bibfnamefont {C.}~\bibnamefont {Zhou}}, \ and\
  \bibinfo {author} {\bibfnamefont {J.}~\bibnamefont {Kurths}},\ }\href@noop {}
  {\bibfield  {journal} {\bibinfo  {journal} {Physical Review E}\ }\textbf
  {\bibinfo {volume} {71}},\ \bibinfo {pages} {016116} (\bibinfo {year}
  {2005})}\BibitemShut {NoStop}%
\bibitem [{\citenamefont {Oh}\ \emph {et~al.}(2005)\citenamefont {Oh},
  \citenamefont {Rho}, \citenamefont {Hong},\ and\ \citenamefont
  {Kahng}}]{oh2005modular}%
  \BibitemOpen
  \bibfield  {author} {\bibinfo {author} {\bibfnamefont {E.}~\bibnamefont
  {Oh}}, \bibinfo {author} {\bibfnamefont {K.}~\bibnamefont {Rho}}, \bibinfo
  {author} {\bibfnamefont {H.}~\bibnamefont {Hong}}, \ and\ \bibinfo {author}
  {\bibfnamefont {B.}~\bibnamefont {Kahng}},\ }\href@noop {} {\bibfield
  {journal} {\bibinfo  {journal} {Physical Review E}\ }\textbf {\bibinfo
  {volume} {72}},\ \bibinfo {pages} {047101} (\bibinfo {year}
  {2005})}\BibitemShut {NoStop}%
\bibitem [{\citenamefont {Belkin}\ and\ \citenamefont
  {Niyogi}(2001)}]{belkin2001laplacian}%
  \BibitemOpen
  \bibfield  {author} {\bibinfo {author} {\bibfnamefont {M.}~\bibnamefont
  {Belkin}}\ and\ \bibinfo {author} {\bibfnamefont {P.}~\bibnamefont
  {Niyogi}},\ }in\ \href@noop {} {\emph {\bibinfo {booktitle} {NIPS}}},\
  Vol.~\bibinfo {volume} {14}\ (\bibinfo {year} {2001})\ pp.\ \bibinfo {pages}
  {585--591}\BibitemShut {NoStop}%
\bibitem [{\citenamefont {Olfati-Saber}\ and\ \citenamefont
  {Murray}(2004)}]{olfati2004consensus}%
  \BibitemOpen
  \bibfield  {author} {\bibinfo {author} {\bibfnamefont {R.}~\bibnamefont
  {Olfati-Saber}}\ and\ \bibinfo {author} {\bibfnamefont {R.~M.}\ \bibnamefont
  {Murray}},\ }\href@noop {} {\bibfield  {journal} {\bibinfo  {journal} {IEEE
  Transactions on automatic control}\ }\textbf {\bibinfo {volume} {49}},\
  \bibinfo {pages} {1520} (\bibinfo {year} {2004})}\BibitemShut {NoStop}%
\bibitem [{\citenamefont {Liljeros}\ \emph {et~al.}(2001)\citenamefont
  {Liljeros}, \citenamefont {Edling}, \citenamefont {Amaral}, \citenamefont
  {Stanley},\ and\ \citenamefont {{\AA}berg}}]{liljeros2001web}%
  \BibitemOpen
  \bibfield  {author} {\bibinfo {author} {\bibfnamefont {F.}~\bibnamefont
  {Liljeros}}, \bibinfo {author} {\bibfnamefont {C.~R.}\ \bibnamefont
  {Edling}}, \bibinfo {author} {\bibfnamefont {L.~A.~N.}\ \bibnamefont
  {Amaral}}, \bibinfo {author} {\bibfnamefont {H.~E.}\ \bibnamefont {Stanley}},
  \ and\ \bibinfo {author} {\bibfnamefont {Y.}~\bibnamefont {{\AA}berg}},\
  }\href@noop {} {\bibfield  {journal} {\bibinfo  {journal} {Nature}\ }\textbf
  {\bibinfo {volume} {411}},\ \bibinfo {pages} {907} (\bibinfo {year}
  {2001})}\BibitemShut {NoStop}%
\bibitem [{\citenamefont {Noh}\ and\ \citenamefont
  {Rieger}(2004)}]{noh2004random}%
  \BibitemOpen
  \bibfield  {author} {\bibinfo {author} {\bibfnamefont {J.~D.}\ \bibnamefont
  {Noh}}\ and\ \bibinfo {author} {\bibfnamefont {H.}~\bibnamefont {Rieger}},\
  }\href@noop {} {\bibfield  {journal} {\bibinfo  {journal} {Physical Review
  Letters}\ }\textbf {\bibinfo {volume} {92}},\ \bibinfo {pages} {118701}
  (\bibinfo {year} {2004})}\BibitemShut {NoStop}%
\bibitem [{\citenamefont {De~Domenico}\ \emph {et~al.}(2014)\citenamefont
  {De~Domenico}, \citenamefont {Sol{\'e}-Ribalta}, \citenamefont {G{\'o}mez},\
  and\ \citenamefont {Arenas}}]{dedomenico2014navigability}%
  \BibitemOpen
  \bibfield  {author} {\bibinfo {author} {\bibfnamefont {M.}~\bibnamefont
  {De~Domenico}}, \bibinfo {author} {\bibfnamefont {A.}~\bibnamefont
  {Sol{\'e}-Ribalta}}, \bibinfo {author} {\bibfnamefont {S.}~\bibnamefont
  {G{\'o}mez}}, \ and\ \bibinfo {author} {\bibfnamefont {A.}~\bibnamefont
  {Arenas}},\ }\href@noop {} {\bibfield  {journal} {\bibinfo  {journal} {PNAS}\
  }\textbf {\bibinfo {volume} {111}},\ \bibinfo {pages} {8351} (\bibinfo {year}
  {2014})}\BibitemShut {NoStop}%
\bibitem [{\citenamefont {De~Domenico}\ \emph {et~al.}(2016)\citenamefont
  {De~Domenico}, \citenamefont {Granell}, \citenamefont {Porter},\ and\
  \citenamefont {Arenas}}]{de2016physics}%
  \BibitemOpen
  \bibfield  {author} {\bibinfo {author} {\bibfnamefont {M.}~\bibnamefont
  {De~Domenico}}, \bibinfo {author} {\bibfnamefont {C.}~\bibnamefont
  {Granell}}, \bibinfo {author} {\bibfnamefont {M.~A.}\ \bibnamefont {Porter}},
  \ and\ \bibinfo {author} {\bibfnamefont {A.}~\bibnamefont {Arenas}},\
  }\href@noop {} {\bibfield  {journal} {\bibinfo  {journal} {Nature Physics}\
  }\textbf {\bibinfo {volume} {12}},\ \bibinfo {pages} {901} (\bibinfo {year}
  {2016})}\BibitemShut {NoStop}%
\bibitem [{\citenamefont {Masuda}\ \emph {et~al.}(2016)\citenamefont {Masuda},
  \citenamefont {Porter},\ and\ \citenamefont {Lambiotte}}]{masuda2016random}%
  \BibitemOpen
  \bibfield  {author} {\bibinfo {author} {\bibfnamefont {N.}~\bibnamefont
  {Masuda}}, \bibinfo {author} {\bibfnamefont {M.~A.}\ \bibnamefont {Porter}},
  \ and\ \bibinfo {author} {\bibfnamefont {R.}~\bibnamefont {Lambiotte}},\
  }\href@noop {} {\bibfield  {journal} {\bibinfo  {journal} {arXiv:1612.03281}\
  } (\bibinfo {year} {2016})}\BibitemShut {NoStop}%
\bibitem [{\citenamefont {Jones}\ \emph {et~al.}(2008)\citenamefont {Jones},
  \citenamefont {Maggioni},\ and\ \citenamefont {Schul}}]{jones2008manifold}%
  \BibitemOpen
  \bibfield  {author} {\bibinfo {author} {\bibfnamefont {P.~W.}\ \bibnamefont
  {Jones}}, \bibinfo {author} {\bibfnamefont {M.}~\bibnamefont {Maggioni}}, \
  and\ \bibinfo {author} {\bibfnamefont {R.}~\bibnamefont {Schul}},\
  }\href@noop {} {\bibfield  {journal} {\bibinfo  {journal} {PNAS}\ }\textbf
  {\bibinfo {volume} {105}},\ \bibinfo {pages} {1803} (\bibinfo {year}
  {2008})}\BibitemShut {NoStop}%
\bibitem [{\citenamefont {Nadler}\ \emph {et~al.}(2008)\citenamefont {Nadler},
  \citenamefont {Lafon}, \citenamefont {Coifman},\ and\ \citenamefont
  {Kevrekidis}}]{nadler2008diffusion}%
  \BibitemOpen
  \bibfield  {author} {\bibinfo {author} {\bibfnamefont {B.}~\bibnamefont
  {Nadler}}, \bibinfo {author} {\bibfnamefont {S.}~\bibnamefont {Lafon}},
  \bibinfo {author} {\bibfnamefont {R.}~\bibnamefont {Coifman}}, \ and\
  \bibinfo {author} {\bibfnamefont {I.~G.}\ \bibnamefont {Kevrekidis}},\ }in\
  \href@noop {} {\emph {\bibinfo {booktitle} {Principal manifolds for data
  visualization and dimension reduction}}}\ (\bibinfo  {publisher} {Springer},\
  \bibinfo {year} {2008})\ pp.\ \bibinfo {pages} {238--260}\BibitemShut
  {NoStop}%
\bibitem [{\citenamefont {Girvan}\ and\ \citenamefont
  {Newman}(2002)}]{girvan2002community}%
  \BibitemOpen
  \bibfield  {author} {\bibinfo {author} {\bibfnamefont {M.}~\bibnamefont
  {Girvan}}\ and\ \bibinfo {author} {\bibfnamefont {M.~E.}\ \bibnamefont
  {Newman}},\ }\href@noop {} {\bibfield  {journal} {\bibinfo  {journal} {PNAS}\
  }\textbf {\bibinfo {volume} {99}},\ \bibinfo {pages} {7821} (\bibinfo {year}
  {2002})}\BibitemShut {NoStop}%
\bibitem [{\citenamefont {Reichardt}\ and\ \citenamefont
  {Bornholdt}(2004)}]{reichardt2004detecting}%
  \BibitemOpen
  \bibfield  {author} {\bibinfo {author} {\bibfnamefont {J.}~\bibnamefont
  {Reichardt}}\ and\ \bibinfo {author} {\bibfnamefont {S.}~\bibnamefont
  {Bornholdt}},\ }\href@noop {} {\bibfield  {journal} {\bibinfo  {journal}
  {Physical Review Letters}\ }\textbf {\bibinfo {volume} {93}},\ \bibinfo
  {pages} {218701} (\bibinfo {year} {2004})}\BibitemShut {NoStop}%
\bibitem [{\citenamefont {Blondel}\ \emph {et~al.}(2008)\citenamefont
  {Blondel}, \citenamefont {Guillaume}, \citenamefont {Lambiotte},\ and\
  \citenamefont {Lefebvre}}]{blondel2008fast}%
  \BibitemOpen
  \bibfield  {author} {\bibinfo {author} {\bibfnamefont {V.~D.}\ \bibnamefont
  {Blondel}}, \bibinfo {author} {\bibfnamefont {J.-L.}\ \bibnamefont
  {Guillaume}}, \bibinfo {author} {\bibfnamefont {R.}~\bibnamefont
  {Lambiotte}}, \ and\ \bibinfo {author} {\bibfnamefont {E.}~\bibnamefont
  {Lefebvre}},\ }\href@noop {} {\bibfield  {journal} {\bibinfo  {journal}
  {Journal of statistical mechanics}\ }\textbf {\bibinfo {volume} {2008}},\
  \bibinfo {pages} {P10008} (\bibinfo {year} {2008})}\BibitemShut {NoStop}%
\bibitem [{\citenamefont {Rosvall}\ and\ \citenamefont
  {Bergstrom}(2007)}]{rosvall2007information}%
  \BibitemOpen
  \bibfield  {author} {\bibinfo {author} {\bibfnamefont {M.}~\bibnamefont
  {Rosvall}}\ and\ \bibinfo {author} {\bibfnamefont {C.~T.}\ \bibnamefont
  {Bergstrom}},\ }\href@noop {} {\bibfield  {journal} {\bibinfo  {journal}
  {PNAS}\ }\textbf {\bibinfo {volume} {104}},\ \bibinfo {pages} {7327}
  (\bibinfo {year} {2007})}\BibitemShut {NoStop}%
\bibitem [{\citenamefont {N{\'e}gyessy}\ \emph {et~al.}(2006)\citenamefont
  {N{\'e}gyessy}, \citenamefont {Nepusz}, \citenamefont {Kocsis},\ and\
  \citenamefont {Bazs{\'o}}}]{negyessy2006prediction}%
  \BibitemOpen
  \bibfield  {author} {\bibinfo {author} {\bibfnamefont {L.}~\bibnamefont
  {N{\'e}gyessy}}, \bibinfo {author} {\bibfnamefont {T.}~\bibnamefont
  {Nepusz}}, \bibinfo {author} {\bibfnamefont {L.}~\bibnamefont {Kocsis}}, \
  and\ \bibinfo {author} {\bibfnamefont {F.}~\bibnamefont {Bazs{\'o}}},\
  }\href@noop {} {\bibfield  {journal} {\bibinfo  {journal} {European Journal
  of Neuroscience}\ }\textbf {\bibinfo {volume} {23}},\ \bibinfo {pages} {1919}
  (\bibinfo {year} {2006})}\BibitemShut {NoStop}%
\bibitem [{\citenamefont {Honey}\ \emph {et~al.}(2007)\citenamefont {Honey},
  \citenamefont {K{\"o}tter}, \citenamefont {Breakspear},\ and\ \citenamefont
  {Sporns}}]{honey2007network}%
  \BibitemOpen
  \bibfield  {author} {\bibinfo {author} {\bibfnamefont {C.~J.}\ \bibnamefont
  {Honey}}, \bibinfo {author} {\bibfnamefont {R.}~\bibnamefont {K{\"o}tter}},
  \bibinfo {author} {\bibfnamefont {M.}~\bibnamefont {Breakspear}}, \ and\
  \bibinfo {author} {\bibfnamefont {O.}~\bibnamefont {Sporns}},\ }\href@noop {}
  {\bibfield  {journal} {\bibinfo  {journal} {PNAS}\ }\textbf {\bibinfo
  {volume} {104}},\ \bibinfo {pages} {10240} (\bibinfo {year}
  {2007})}\BibitemShut {NoStop}%
\bibitem [{\citenamefont {Serrano}\ \emph {et~al.}(2008)\citenamefont
  {Serrano}, \citenamefont {Krioukov},\ and\ \citenamefont
  {Bogun{\'a}}}]{serrano2008self}%
  \BibitemOpen
  \bibfield  {author} {\bibinfo {author} {\bibfnamefont {M.~A.}\ \bibnamefont
  {Serrano}}, \bibinfo {author} {\bibfnamefont {D.}~\bibnamefont {Krioukov}}, \
  and\ \bibinfo {author} {\bibfnamefont {M.}~\bibnamefont {Bogun{\'a}}},\
  }\href@noop {} {\bibfield  {journal} {\bibinfo  {journal} {Physical Review
  Letters}\ }\textbf {\bibinfo {volume} {100}},\ \bibinfo {pages} {078701}
  (\bibinfo {year} {2008})}\BibitemShut {NoStop}%
\bibitem [{\citenamefont {Boguna}\ \emph {et~al.}(2009)\citenamefont {Boguna},
  \citenamefont {Krioukov},\ and\ \citenamefont
  {Claffy}}]{boguna2009navigability}%
  \BibitemOpen
  \bibfield  {author} {\bibinfo {author} {\bibfnamefont {M.}~\bibnamefont
  {Boguna}}, \bibinfo {author} {\bibfnamefont {D.}~\bibnamefont {Krioukov}}, \
  and\ \bibinfo {author} {\bibfnamefont {K.~C.}\ \bibnamefont {Claffy}},\
  }\href@noop {} {\bibfield  {journal} {\bibinfo  {journal} {Nature Physics}\
  }\textbf {\bibinfo {volume} {5}},\ \bibinfo {pages} {74} (\bibinfo {year}
  {2009})}\BibitemShut {NoStop}%
\bibitem [{\citenamefont {Papadopoulos}\ \emph {et~al.}(2012)\citenamefont
  {Papadopoulos}, \citenamefont {Kitsak}, \citenamefont {Serrano},
  \citenamefont {Bogun{\'a}},\ and\ \citenamefont
  {Krioukov}}]{papadopoulos2012popularity}%
  \BibitemOpen
  \bibfield  {author} {\bibinfo {author} {\bibfnamefont {F.}~\bibnamefont
  {Papadopoulos}}, \bibinfo {author} {\bibfnamefont {M.}~\bibnamefont
  {Kitsak}}, \bibinfo {author} {\bibfnamefont {M.~{\'A}.}\ \bibnamefont
  {Serrano}}, \bibinfo {author} {\bibfnamefont {M.}~\bibnamefont {Bogun{\'a}}},
  \ and\ \bibinfo {author} {\bibfnamefont {D.}~\bibnamefont {Krioukov}},\
  }\href@noop {} {\bibfield  {journal} {\bibinfo  {journal} {Nature}\ }\textbf
  {\bibinfo {volume} {489}},\ \bibinfo {pages} {537} (\bibinfo {year}
  {2012})}\BibitemShut {NoStop}%
\bibitem [{\citenamefont {Kleineberg}\ \emph {et~al.}(2016)\citenamefont
  {Kleineberg}, \citenamefont {Bogu{\~n}{\'a}}, \citenamefont
  {{\'A}ngeles~Serrano},\ and\ \citenamefont
  {Papadopoulos}}]{kleineberg2016hidden}%
  \BibitemOpen
  \bibfield  {author} {\bibinfo {author} {\bibfnamefont {K.-K.}\ \bibnamefont
  {Kleineberg}}, \bibinfo {author} {\bibfnamefont {M.}~\bibnamefont
  {Bogu{\~n}{\'a}}}, \bibinfo {author} {\bibfnamefont {M.}~\bibnamefont
  {{\'A}ngeles~Serrano}}, \ and\ \bibinfo {author} {\bibfnamefont
  {F.}~\bibnamefont {Papadopoulos}},\ }\href@noop {} {\bibfield  {journal}
  {\bibinfo  {journal} {Nature Physics}\ }\textbf {\bibinfo {volume} {12}},\
  \bibinfo {pages} {1076} (\bibinfo {year} {2016})}\BibitemShut {NoStop}%
\bibitem [{\citenamefont {De~Domenico}\ \emph {et~al.}(2013)\citenamefont
  {De~Domenico}, \citenamefont {Sol{\'e}-Ribalta}, \citenamefont {Cozzo},
  \citenamefont {Kivel{\"a}}, \citenamefont {Moreno}, \citenamefont {Porter},
  \citenamefont {G{\'o}mez},\ and\ \citenamefont
  {Arenas}}]{de2013mathematical}%
  \BibitemOpen
  \bibfield  {author} {\bibinfo {author} {\bibfnamefont {M.}~\bibnamefont
  {De~Domenico}}, \bibinfo {author} {\bibfnamefont {A.}~\bibnamefont
  {Sol{\'e}-Ribalta}}, \bibinfo {author} {\bibfnamefont {E.}~\bibnamefont
  {Cozzo}}, \bibinfo {author} {\bibfnamefont {M.}~\bibnamefont {Kivel{\"a}}},
  \bibinfo {author} {\bibfnamefont {Y.}~\bibnamefont {Moreno}}, \bibinfo
  {author} {\bibfnamefont {M.~A.}\ \bibnamefont {Porter}}, \bibinfo {author}
  {\bibfnamefont {S.}~\bibnamefont {G{\'o}mez}}, \ and\ \bibinfo {author}
  {\bibfnamefont {A.}~\bibnamefont {Arenas}},\ }\href@noop {} {\bibfield
  {journal} {\bibinfo  {journal} {Physical Review X}\ }\textbf {\bibinfo
  {volume} {3}},\ \bibinfo {pages} {041022} (\bibinfo {year}
  {2013})}\BibitemShut {NoStop}%
\bibitem [{\citenamefont {Del~Genio}\ \emph {et~al.}(2016)\citenamefont
  {Del~Genio}, \citenamefont {G{\'o}mez-Garde{\~n}es}, \citenamefont
  {Bonamassa},\ and\ \citenamefont {Boccaletti}}]{del2016synchronization}%
  \BibitemOpen
  \bibfield  {author} {\bibinfo {author} {\bibfnamefont {C.~I.}\ \bibnamefont
  {Del~Genio}}, \bibinfo {author} {\bibfnamefont {J.}~\bibnamefont
  {G{\'o}mez-Garde{\~n}es}}, \bibinfo {author} {\bibfnamefont {I.}~\bibnamefont
  {Bonamassa}}, \ and\ \bibinfo {author} {\bibfnamefont {S.}~\bibnamefont
  {Boccaletti}},\ }\href@noop {} {\bibfield  {journal} {\bibinfo  {journal}
  {Science Advances}\ }\textbf {\bibinfo {volume} {2}},\ \bibinfo {pages}
  {e1601679} (\bibinfo {year} {2016})}\BibitemShut {NoStop}%
\end{thebibliography}%


\renewcommand*{\thefigure}{{\bf \arabic{figure}}}
\renewcommand{\figurename}{{\bf Supplementary Figure}}
\renewcommand*{\thetable}{{\bf \arabic{table}}}
\renewcommand{\tablename}{{\bf Supplementary Table}}

\clearpage
\begin{figure*}[!h]
\centering
\includegraphics[width=16cm]{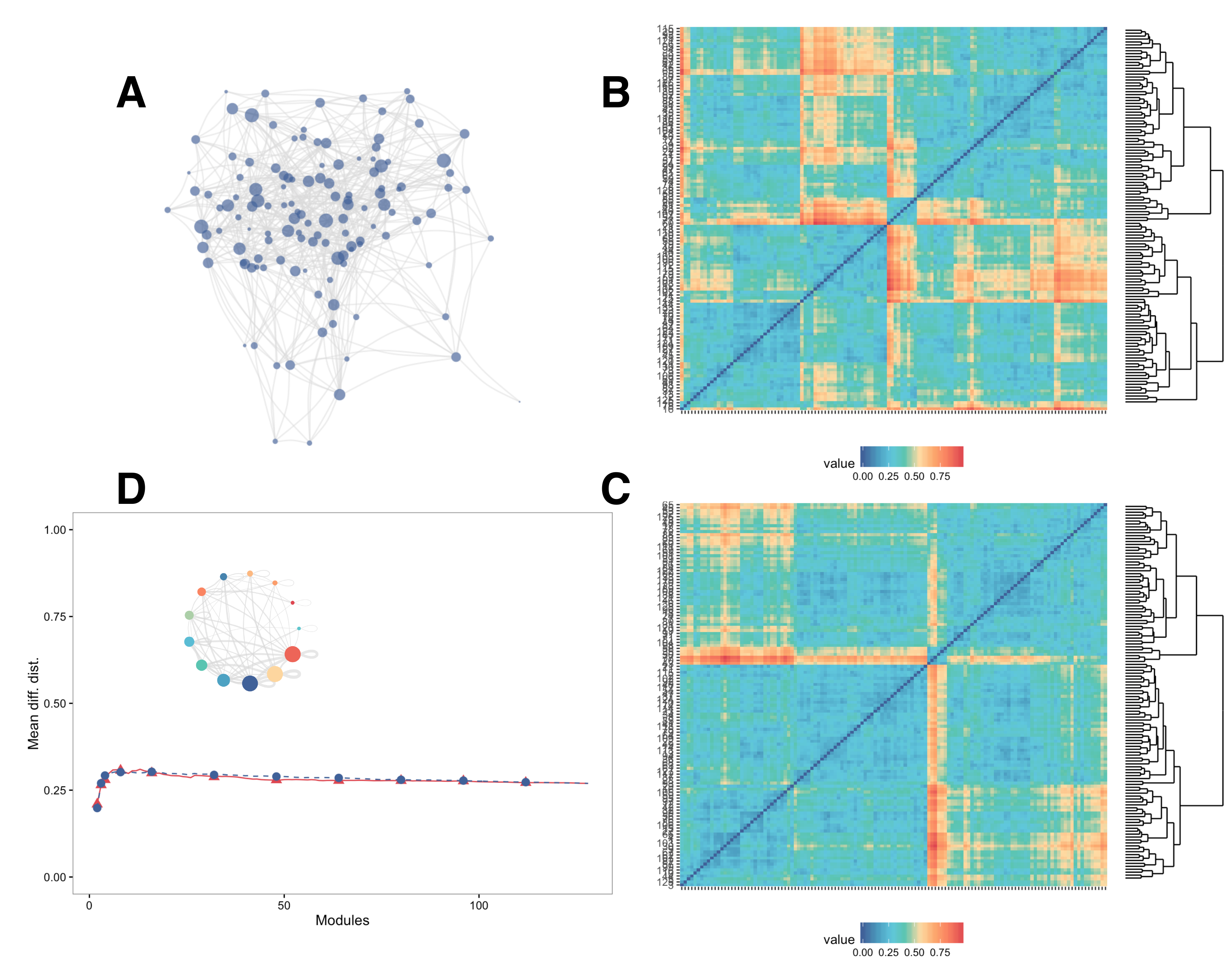}
\caption{\label{fig1}\small{\textbf{Functional mesoscale organization in an Erdos-Renyi network.} (A) An Erdos-Renyi network is not expected to show peculiar mesoscale functional organization (B) when compared to its configuration model (C). Here, diffusion distance matrices are shown in both cases, with color encoding the diffusion distance. In (D), the average diffusion distance in the network of super-units is used to find the most persistent mesoscale in the original network (solid line) and to evaluate its difference from the configuration model (dashed line). The two curves collapse on each other across all scales, correctly suggesting that the identified mesoscale is compatible with its random expectation. Networks with $N=128$ nodes and $p=0.076$ have been considered in this case.}}
\end{figure*}

\newpage
\begin{figure*}[!h]
\centering
\includegraphics[width=16cm]{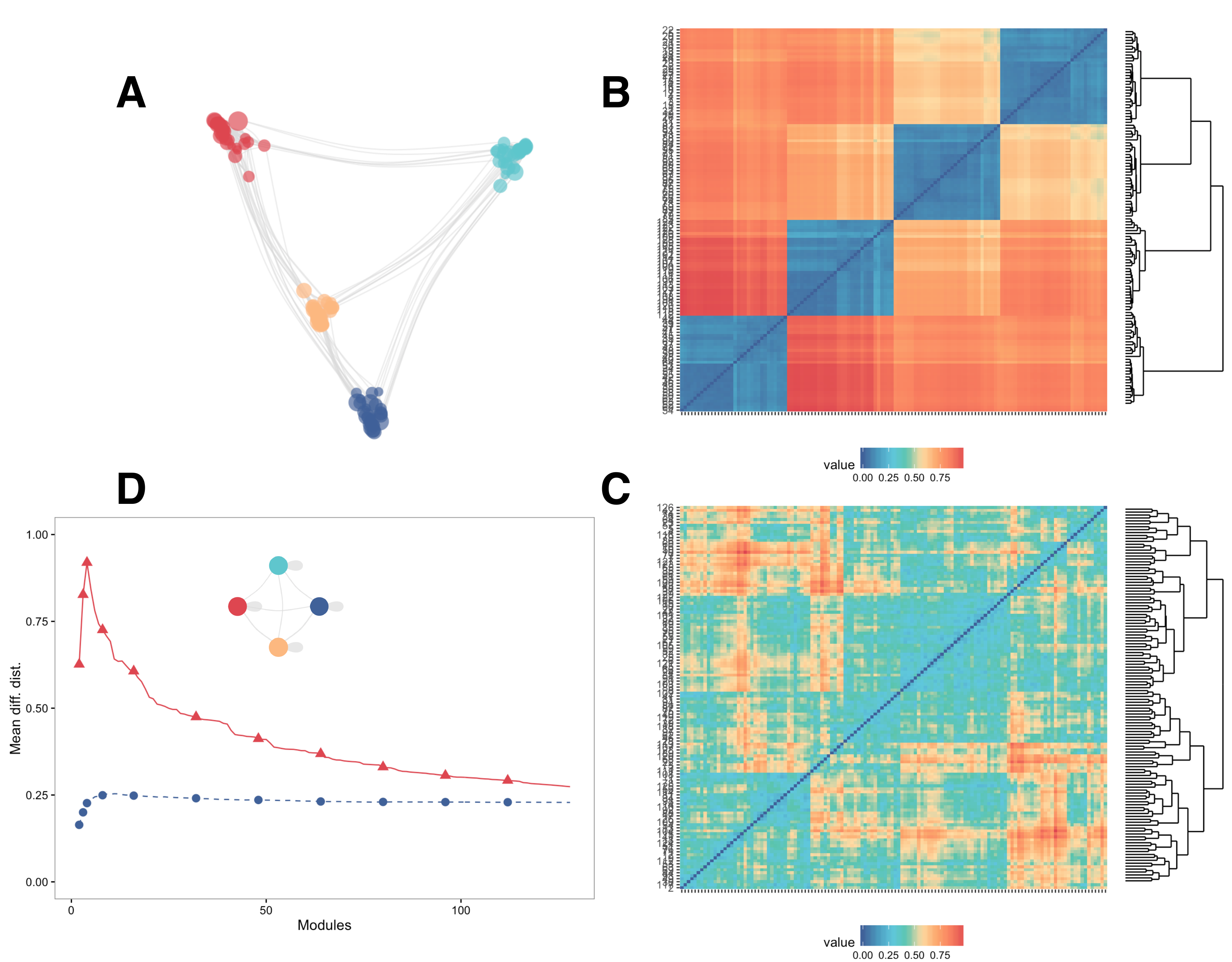}
\caption{\label{fig2}\small{\textbf{Functional mesoscale organization in a Girvan-Newman network.} As in Fig.~\ref{fig1}. In this case there is a strong topological mesoscale structure with 4 clusters and the functional mesoscale consists of the same clusters, providing evidence that functional organization might correspond to structural organization. The difference between the original network and its random expectation is evident in (D). Networks with $N=128$ nodes have been considered in this case.}}
\end{figure*}

\newpage
\begin{figure*}[!t]
\centering
\includegraphics[width=16cm]{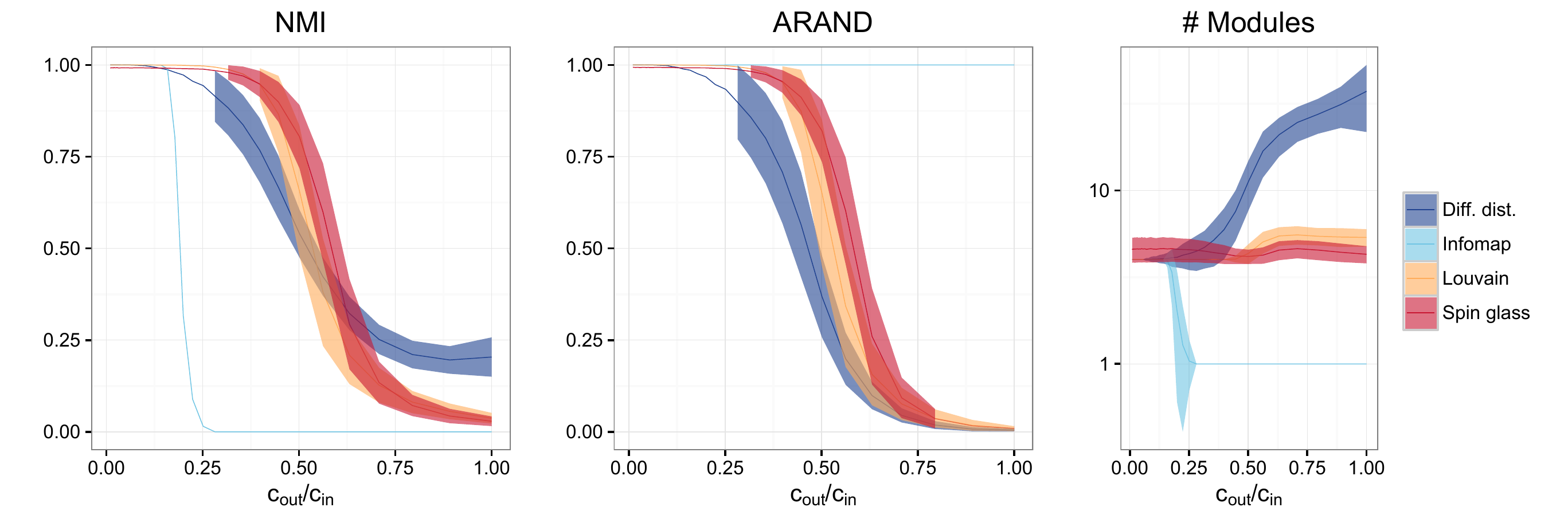}
\caption{\label{fig:bench}\small{\textbf{Functional versus mesoscale organization in Girvan-Newman benchmarks~\cite{girvan2002community}.} We have generated ensembles of random realizations by varying the ratio between the probability of inter-community links ($c_{out}$) and intra-community links ($c_{in}$) between $10^{-3}$ (strong structural mesoscale organization) to 1 (no mesoscale). Methods for detection of structural clusters (Louvain~\cite{blondel2008fast} and Spin Glass~\cite{reichardt2004detecting}) are compared to a method based on compression of information flow (Infomap~\cite{rosvall2007information,rosvall2008maps}) and to functional clusters revealed by diffusion geometry. Two evaluation scores, to compare the revealed clusters against the ground truth, are used: Normalized Mutual Information (NMI, left panel) and Adjusted Rand Index (ARAND, middle panel). The number of clusters is shown in the right panel. The average value (solid lines) and standard deviation (shaded area) are reported.}}
\end{figure*}

\begin{figure*}[!t]
\centering
\includegraphics[width=16cm]{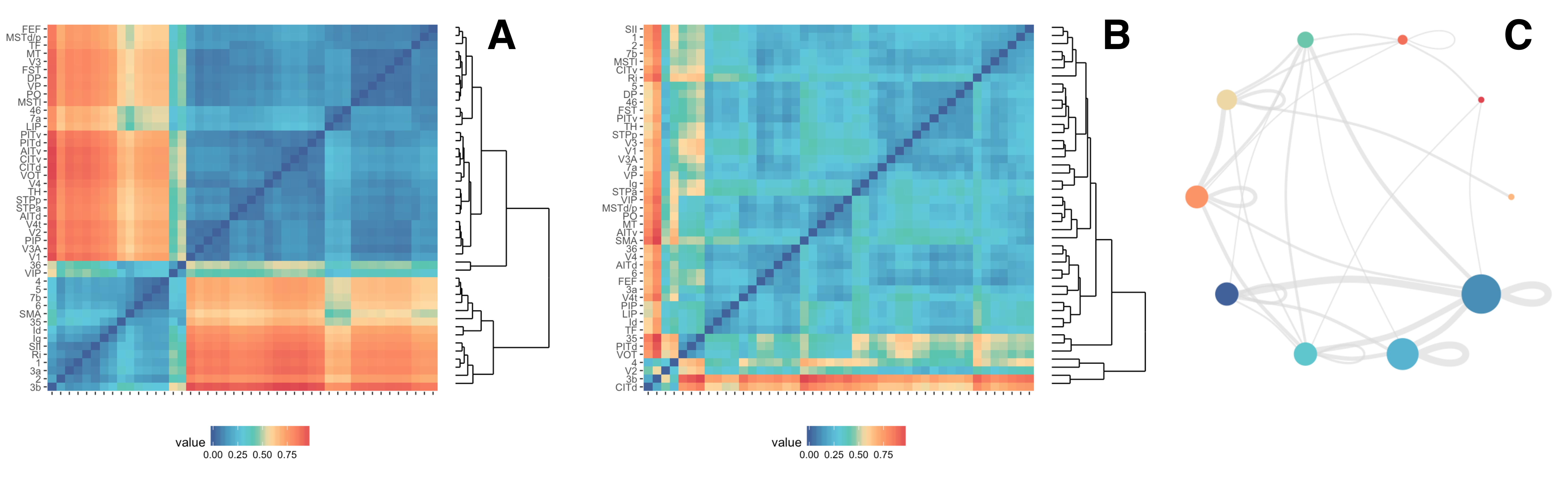}
\caption{\label{fig:app}\small{\textbf{Functional clusters in Macaque visuo-tactile cortex.} Diffusion geometry analysis of the anatomical connectivity (335 visual, 85 sensorimotor and 43 heteromodal) from 30 visual cortical areas and 15 sensorimotor areas in the Macaque monkey~\cite{negyessy2006prediction} clearly reveals the regions corresponding to the two
areas, while unraveling more detailed functional clusters which are persistent across time. (A) Average diffusion distance matrix of the empirical network, (B) of its configuration model and (C) network of functional super-units which is most persistent across time and significantly different from random expectation.}}
\end{figure*}

\begin{figure*}[!t]
\centering
\includegraphics[width=16cm]{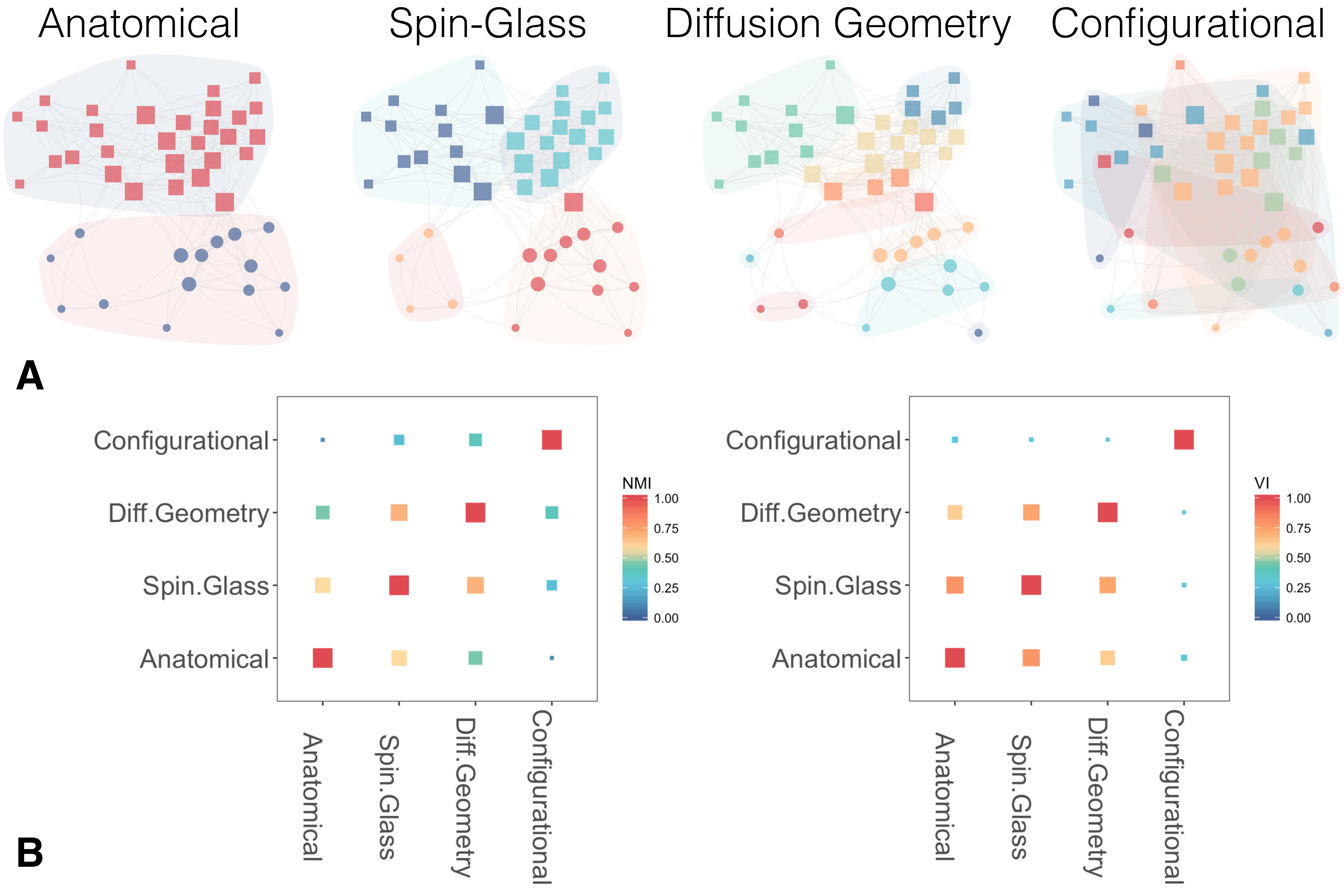}
\caption{\label{fig:comp}\small{\textbf{Novel information from diffusion geometry.} The network representation of the Macaque visuo-tactile cortex is shown in (A). Node's shape encodes anatomical information (circles for sensorimotor areas, squares for visual ones). From left to right, nodes' and groups' color encode the clusters identified from anatomical information only, spin-glass community detection~\cite{reichardt2004detecting}, diffusion geometry functional clusters and configurational clusters (i.e., obtained from a representative random realization of the empirical network, while preserving the underlying degree distribution). (B) The similarity among the identified clusters is quantified by normalized mutual information (left) and variation of information (right).}}
\end{figure*}

\end{document}